\documentstyle[prl,aps,multicol,amssymb]{revtex}
\begin{document}

%
\title{\boldmath Propylene Carbonate Reexamined:\\
       Mode-Coupling $\beta$ Scaling without Factorisation ?}

\author{J.~Wuttke$^1$, M.~Ohl$^2$, M.~Goldammer$^1$, S.~Roth$^1$,
        U.~Schneider$^2$, P.~Lunkenheimer$^2$, R.~Kahn$^3$, 
        B.~Ruffl\'e$^4$, R.~Lechner$^4$, M.~A.~Berg$^5$}
\address{$^1$ Physik-Department E13, Technische Universit\"at M\"unchen, 
         85747 Garching, Germany}
\address{$^2$ Experimentalphysik V, 
         Universit\"at Augsburg, 86135 Augsburg, Germany}
\address{$^3$ Laboratoire L\'eon Brillouin,
         Centre d'\'etudes nucl\'eaires de Saclay, 91191 Gif-sur-Yvette,
         France}   
\address{$^4$ BENSC, Hahn-Meitner-Institut,
         14109 Berlin, Germany}
\address{$^5$ Department of Chemistry and Biochemistry, 
         University of South Carolina, Columbia, SC 29208, USA}

\date{submitted to Phys.~Rev.~E, \today.}

\maketitle

\begin{abstract}
The dynamic susceptibility of propylene carbonate in the 
moderately viscous regime above~$T_{\rm c}$ is reinvestigated 
by incoherent neutron and depolarised light scattering,
and compared to dielectric loss and solvation response. 
Depending on the strength of $\alpha$ relaxation,
a more or less extended $\beta$ scaling regime is found.
Mode-coupling fits yield consistently $\lambda=0.72$ and $T_{\rm c}=182\,$K,
although different positions of the susceptibility minimum indicate
that not all observables have reached the universal asymptotics.
\end{abstract}

\begin{multicols}{2}

\section{Introduction}

\subsection{Motivation}

The glass transition, an essentially dynamic phenomenon, 
can be described as the slowing down and eventual freezing of 
$\alpha$ relaxation.
According to mode-coupling (MC) theory \cite{Got91,x15},
the physical origin of this process must be sought on a picosecond scale
where long-ranged transport starts to evolve 
from vibrational short-time dynamics.

In the long-time limit, MC theory reproduces the well-established
phenomenology of $\alpha$ relaxation.
New results are obtained for shorter times.
In particular, the theory predicts a change of transport mechanism 
around a cross-over temperature~$T_{\rm c}$, 
located in the moderately viscous liquid phase well above the
conventional glass transition temperature~$T_g$.
On cooling towards $T_{\rm c}$, particles spend more and more time
being trapped in transient cages;
this process, labelled fast $\beta$ relaxation, 
is predicted to obey remarkably universal scaling.

In a couple of structural glass formers,
MC predictions have been confirmed \cite{GoSj92,Got99}
primarily by different scattering techniques \cite{CuLD95,PeWu95}.
More recently, the GHz--THz dynamics also became accessible 
by dielectric spectroscopy \cite{LuPD96b,LuPL97}.
Although results are in accord with the MC scenario,
in several cases the data did not fall together
with dynamic susceptibilities from scattering experiments \cite{LuPD97b,ScLB98}
which is in conflict with the asymptotic factorisation property
of MC $\beta$ relaxation.
Recent theoretical developments suggest explanations on the basis of
corrections to scaling \cite{FrFG97b,FuGM98}
and orientational degrees of freedom \cite{ScSc97,FrFG97c,FaSS98}.
For experimental tests, 
more detailed comparisons of different observables are needed.

In this context, 
we performed incoherent neutron and depolarised light scattering experiments
on propylene carbonate (PC, 4-methyl-1,3-dioxolan-2-on, C$_4$O$_2$H$_6$),
a fragile glass former ($T_g=160\,$K) with low molecular weight ($M=102.1$)
which has already been studied by various experimental techniques.
A synopsis of available data shall be given in subsection I.C
after collecting the essential results of MC theory in I.B.

\subsection{The Mode-Coupling Cross-Over}

In its simplest (``idealised'') formulation, 
MC theory describes an ergodic-to-nonergodic transition at $T_{\rm c}$.
On the low-temperature side,
the onset of fast $\beta$ relaxation leads 
to an anomalous decrease of the Debye-Waller or Lamb-M\"o\ss bauer factor,
\begin{equation}\label{Esqrt}
   f_q = f_q^c + h_q |\sigma|^{1/2} {\,,~}\sigma>0\,,
\end{equation}
with a reduced temperature $\sigma=(T_{\rm c}-T)/T_{\rm c}$.
On the high-temperature side,
the time constant of $\alpha$ relaxation diverges 
with a fractal exponent~$\gamma$,
\begin{equation}\label{Etaua}
   \tau \propto |\sigma|^{-\gamma} {\,,~}\sigma<0\,.
\end{equation}

Such a transition has actually been observed in a colloidal suspension
\cite{Meg95}.
In a structural glass former, 
the singularities
(\ref{Esqrt}) and~(\ref{Etaua}) are smeared out
because activated hopping processes restore ergodicity \cite{GoSj87b}.
Under these limitations, 
integral quantity like $f_q$ or~$\tau_\alpha$
do not allow for decisive tests of theory.
One rather has to study the full dynamics,
as represented by the dynamic susceptibility $\chi''_q(\omega)$,
or any other dynamic variable coupling to it.

Stronger predictions are made for the fast $\beta$ regime:
around the minimum between $\alpha$ peak and vibrational excitations,
in a temperature range close enough to, but sufficiently above $T_{\rm c}$,
any susceptibility is expected to reach the same asymptotic limit
\begin{equation}\label{EminL}
   \chi''(\omega) = \chi_\sigma g_\lambda(\omega/\omega_\sigma)
\end{equation}
where the scaling function $g_\lambda$ is fully determined by 
one single parameter~$\lambda$ \cite{Got90}.
Amplitude and frequency scale of (\ref{EminL}) should become singular
on cooling towards $T_{\rm c}$,
\begin{equation}\label{Epars}
    \chi_\sigma \propto |\sigma|^{1/2}\,{\rm,~and~} 
   \omega_\sigma \propto |\sigma|^{1/2a} {\,,~}\sigma<0
\end{equation}
where the exponent~$a$, just as $\gamma$ in (\ref{Etaua}),
 is determined by~$\lambda$.

\subsection{Previous Studies of Propylene Carbonate}

Table~1 summarises previous studies of fast dynamics in PC.
All authors reported at least partial accord with MC predictions.
However, 
not all MC interpretations were consistent with each other.
It is an essential result of idealised MC theory
that in the asymptotic $\beta$ regime all 
dynamic observables show the same spectral distribution and 
the same temperature dependence.
Therefore, 
one material is characterised by just one~$\lambda$ and one~$T_{\rm c}$.
Some of the $T_{\rm c}$ reported for PC 
are therefore incompatible with MC theory.

As with other materials,
early scattering experiments \cite{BoHo91,ElBT92} concentrated 
on the square-root singularity of~$f_q$.
As in other materials, 
this singularity remains elusive:
in Brillouin scattering, 
data fitting depends on uncontrolled approximations 
for the memory function \cite{DuLC94}.
Similarly,
in neutron scattering a determination of $T_{\rm c}$ from (\ref{Esqrt})
works at best if the full lineshape on the $\sigma>0$ side
is known from high-resolution spectroscopy \cite{BaFK89b}.
Therefore, the isolated results $T_{\rm c}=210$ or 270\,K
can be discarded from further consideration.

In comparison, 
the determination of $T_{\rm c}$ from viscosity or $\alpha$-relaxation data
works better.
However,
available data do not allow for an independent determination of~$\lambda$,
and even for one given value $\lambda=0.70$, results vary
between $T_{\rm c}=180$ and 196\,K.

A dynamic susceptibility has been measured first 
by depolarised light scattering \cite{DuLC94},
yielding $\lambda=0.78$ and $T_{\rm c}=187$.
These values have shown to be consistent with
dielectric loss spectroscopy \cite{LuPD97b,LuPD96c,ScLB99}.
However, a time-domain optical measurement of the solvation response 
of a solute molecule found a significantly lower~$T_{\rm c}$ 
if the light scattering value of $\lambda=0.78$ was assumed 
\cite{MaBB95,MaBB96}.

In the following section II, 
we study the fast relaxation regime by neutron scattering.
The $\lambda$ and $T_{\rm c}$ we obtain 
differ significantly from results of the other dynamic measurements
which motivates us to remeasure some light scattering spectra (sect.~III) 
and to reanalyse dielectric-loss and solvation-response data (sect.~IV).
Only then, all susceptibilities will be compared in section~V.

\section{Neutron Scattering}

\subsection{Experiments}

Inelastic, incoherent neutron scattering has been measured 
on the time-of-flight spectrometers Mib\'emol at the 
Laboratoire L\'eon Brillouin, Saclay,
and {\sc Neat} at the Hahn-Meitner-Institut, Berlin.

On Mib\'emol, the counter-rotating choppers were operated with 10\,000~rpm.
With an incident neutron wavelength of $\lambda_{\rm i}=8.5\,$\AA,
we obtained a resolution (fwhm) of $6.5-9.5\,$GHz, depending on angle.
At the Berlin reactor, the flux delivered by the undermoderated cold source 
decreases rather fast for wavelengths beyond 5 or 6\,\AA;
therefore $\lambda_{\rm i}=5.5\,$\AA\ was chosen.
With counter-rotating choppers at 20\,000~rpm,
we achieved nevertheless a resolution of $11-13.5\,$GHz.
Total count rates were of the same order for both instruments;
a precise comparison cannot be made 
because different sample geometries were used.

For the Mib\'emol experiment,
we used an Al hollow cylinder \cite{Wut99} 
with 30\,mm outer diameter and
a sample layer of about 0.1\,mm thickness.
In this container,
the sample crystallised partially during a 190\,K scan;
at 175\,K, 
as far as one can tell from the elastic structure factor of a 
predominantly incoherent scatterer,
complete or nearly complete crystallisation occured within 2 hours.
On the other hand, after rapidly cooling from 260\,K,
it was possible to perform a 2\,K scan in the fully amorphous state.

On {\sc Neat}, we tried to remeasure the dynamics of the supercooled state
below 200\,K.
To this purpose,
we filled the sample into nearly 200 thin capillaries 
(soda lime, inner diameter 0.2\,mm, Hilgenberg).
As in other liquids, 
this packaging proved highly successful;
comparison to a vanadium scattering and visual inspection showed that
down to 168\,K no crystallisation occured.
Because the detectors are located at only 250\,cm from the sample,
the resolution of {\sc Neat} is very sensitive to flightpath differences.
We therefore abstained from using a hollow cylindrical geometry;
instead, we placed the capillaries in a rectangular holder 
with a 30\,mm base length
which was then mounted at $45^\circ$ with respect to the incoming beam.
For both experiments,
the sample material, 1,2-propylene carbonate (99.7\,\%, Sigma--Aldrich),
was loaded under an inert gas.

After converting the raw data to $S(2\theta,\omega)$,
they were binned in about 30 angular groups;
a non-equidistant frequency binning was imposed by 
requiring statistic fluctuations to fall below a given mark.
Only then was the container scattering subtracted.
The absolute intensity scale was taken from the elastic scattering
of the Mib\'emol 2\,K scan;
for {\sc Neat}, detectors were first calibrated to vanadium
before the overall scale was fit to Mib\'emol at 210\,K.

The scattering law $S(2\theta,\omega)$ depends still on $\lambda_{\rm i}$.
Only after interpolation to constant wavenumbers~$q$
one obtains spectra $S(q,\omega)$ 
which are independent of the kinematics of scattering,
and only then a direct comparison between our two experiments becomes possible.
This comparison is performed explicitely at 207/210\,K in Fig.~\ref{Fnsqw}.
In the quasielastic scattering range, up to some 100\,GHz,
the accord between Mib\'emol and {\sc Neat} is excellent.

\subsection{The Factorisation Property}

For analysing the data, especially in the quasielastic range,
it is advantageous to visualise the scattering law as a susceptibility 
\begin{equation}\label{Etriv}
\chi''_q(\omega) = S(q,\omega) / n(\omega)
\end{equation}
with the Bose factor $n(\omega)= {(\exp(\hbar\omega/k_{\rm B}T)-1)}^{-1}$.

Around the susceptibility minimum,
mode coupling theory predicts
that the $\beta$ line shape is asymptotically
the same for all experimental observables
which couple to density fluctuations;
thus {\it a fortiori}\/ 
it must be the same for incoherent neutron susceptibilities
measured at different wavenumbers~$q$.
We therefore expect $\chi''_q(\omega)$ to factorise into 
an $\omega$-independent amplitude
and a $q$-independent spectral function,
\begin{equation}\label{Efact}
   \chi''_q(\omega) = h_q \chi''(\omega)\;.
\end{equation}
On different theoretical grounds,
such a factorisation is also expected for incoherent scattering
from harmonic vibrations in lowest order of mass expansion 
 \cite{Pla54,WuKB93}.

There are several ways to test (\ref{Efact}) and to determine $h_q$:
{\it e.\,g.}, one may iteratively construct a model for $\chi''(\omega)$ 
and fit it to the individual $\chi''_q(\omega)$ data sets.
More simply, $h_q$ can be calculated from a least-squares match between 
neighbouring $q$ cuts \cite{WuKB93}.
For the present data, 
we find that $h_q$ does not depend on the chosen procedure,
nor does it vary with the frequency subrange from which it is determined.

A first surprising result is then 
the strictly linear wavenumber dependence of $h_q$
(Fig.~\ref{Ffact1}a).
Within MC theory, $q$ dependences can be calculated 
only if a specific microscopic structure is put in.
Within harmonic theory, however, 
there is a clear expectation that $h_q\propto q^2$.
Astonishingly enough, 
this $q^2$ dependence is never seen. 
Instead, in at least two other molecular systems
a linear $h_q\propto q$ dependence is found 
though the temperature dependence of THz modes 
indicates pure harmonic behaviour \cite{KiBD92,x34}.
From other time-of-flight studies \cite{WuKB93,CuDo90} we suspect
and a simulation \cite{x41} confirms 
that this is mainly a multiple-scattering effect,
Anyway, the linearity is so accurate 
that in our analysis we shall use (\ref{Efact}) with
$h_q=q/q_0$ (with an arbitrary normalisation $q_0=1\,$\AA$^{-1}$)
instead of employing the empirical values.

Fig.~\ref{Fnsu} shows that
the factorisation holds over the full experimental data range, 
at least from 30 to 2500\,GHz,
except for the onset of $\alpha$ relaxation at the highest temperature
(which is associated with long-ranged transport 
and depends therefore strongly on~$q$).
This allows us to condense our huge two-dimensional data sets into
$q$-independent functions $\chi''(\omega)$ with much improved statistics;
only on this basis is it possible to proceed with a 
quantitative analysis of noisy data accumulated in relatively short scans.

Before proceeding with the analysis
we have to note that the $\chi''(\omega)$ determined by this method
and shown for 207/210\,K in Fig.~\ref{Ffact1}b
reveals considerable discrepancies 
between Mib\'emol and {\sc Neat} 
at frequencies above about 400\,GHz.
Although these deviations passed unnoticed on the double logarithmic scale 
of Fig.~\ref{Fnsqw}b, they are already present in the raw data.
While we can exclude container scattering and dark counts as possible causes,
two other frequency-dependent effects are likely to contribute:
multiple scattering \cite{x41}
and an inaccurate detector efficiency correction \cite{x42}.
Multiple scattering cannot be corrected for
unless a comprehensive theoretical model of $S(q,\omega)$ is used as input;
therefore,
it presents a fundamental limitation to the determination of spectral shapes
by quasielastic neutron scattering.

\subsection{\boldmath Master Curves for $\beta$ Relaxation}

The $q$-independent $\chi''(\omega)$ can now be used
to test the scaling form (\ref{EminL}).
We use fits with a fixed~$\lambda$ to determine
$\chi_\sigma$ and $\omega_\sigma$.
With these values, 
a master curve $\chi''(\omega/\omega_\sigma)/\chi_\sigma$ is constructed,
and from the master curve the scaling range is read off  
which is then used for improved fits to the original data.
Figure~\ref{Fnsufi} shows fits to the original 
 $\chi''(\omega)$ for $\lambda=0.72$,
and Figure~\ref{F3nma} shows master curves for three $\lambda$'s.

In principle, 
a self-consistent $\lambda$ can be determined in an iterative procedure
from free fits to the master curves.
For our $220-260\,$K data, $\lambda$ tends towards values around 0.69
(Figure~\ref{F3nma}a),
but the convergence is erratic, 
and the outcome depends on the subjective decision of 
which points to include in the master curve.

With the data on hand, 
a more restrictive determination of $\lambda$ 
is possible from the temperature dependence~(\ref{Epars}).
The insets in Figure~\ref{F3nma} show that
a consistent linear behaviour 
${\chi_\sigma}^2\propto{\omega_\sigma}^{2a}\propto\sigma$
is found only for $\lambda=0.72$ and $T_{\rm c}=182\,$K; 
the exponent $a=0.32$, which corresponds to $\lambda=0.72$,
is confirmed by cross-checking
$\ln \chi_\sigma$ {\it vs.} $\ln \omega_\sigma$ 
(inset of Figure~\ref{Fnsufi}).

Figure~\ref{F3nma}c demonstrates that the value $\lambda=0.78$
suggested by preceeding light scattering and dielectric loss measurements
do not give a good description of the neutron scattering data:
the master curve is of poorer quality than for $\lambda=0.69$ or 0.72,
especially at frequencies around and below the minimum;
free fits with $g_\lambda$ show a clear trend towards smaller values of
$\lambda$;
and the scales $\chi_\sigma$ and $\omega_\sigma$ 
do not consistently follow~(\ref{Epars}).

This discrepancy motivates us to remeasure some light scattering spectra
around the susceptibility minimum.

\section{Light Scattering}

\subsection{Previous Studies}

Light scattering, just as neutron scattering, 
comprises coherent and incoherent contributions.
Since the wavelength of visible light is much longer than molecular dimensions,
coherent scattering arises only from sound modes,
giving rise to discrete Brillouin lines.
Incoherent scattering, on the other hand, 
sees local motion and yields a continuous spectrum.
In a very first approximation, this spectrum can be interpreted 
as if it were a $q$ averaged neutron scattering law;
microscopic models suggest that the scattering mechanism involves
four-point density correlations and/or rotational motion.

Brillouin scattering
yields the velocity and damping constant of sound waves.
In principle \cite{FuGL90}, 
the sound dispersion through the glass transition reveals 
the strength of the $\alpha$ relaxation so that it is possible
to read off the Debye-Waller factor~$f_q$ from fits to the Brillouin lines.
In practice,
this led to an inconsistent estimate of~$T_{\rm c}$ in PC \cite{ElBT92};
for a reliable determination of $f_q(T)$,
one should not only determine one limiting sound velocity 
by ultrasonic experiments,
but also provide an independently determined memory function
as input \cite{DuLC94}.

More direct information on the microscopic dynamics is obtained from
the incoherent continuous spectra.
In order to suppress the much stronger Brillouin lines, 
these spectra are preferentially gathered
in depolarised (VH) back-scattering geometry.
In an extensive study of PC, 
besides a thorough discussion of Brillouin scattering,
Du {\it et al.} \cite{DuLC94}
have also measured VH spectra over a wide frequency band,
made accessible by combining 
a tandem interferometer with a grating monochromator.
The measurements extended over a large temperature range, 
including $\alpha$ relaxation as well as hopping processes below $T_{\rm c}$,
and were analysed within 
both the ``idealised'' and ``extended'' version of MC theory.

While these measurements gave a broad overview 
of the dynamics of glass-forming PC, 
they did not yield spectral lineshapes with the precision we need now 
for a quantitative comparison with neutron scattering results.
In particular,
the exponent parameter $\lambda=0.78$,
obtained within idealised theory from a global fit to $T>T_{\rm c}$ data,
was given with a relatively large error range of $\pm0.05$.
Furthermore,
as in other broad band light scattering studies performed 
until quite recently,
the tandem interferometer was used in series with an insufficient bandpass
which did not fully suppress higher-order transmissions of the interferometer
\cite{SuWN98,BaLS99}.
For these reasons, we remeasured some depolarised spectra,
concentrating on the temperature and frequency range 
of the asymptotic $\beta$ regime. 

\subsection{Experimental Set-up}

Experiments were performed in Garching on 
a Fabry-Perot six-pass tandem interferometer.
The instrument, bought from J.~R.\ Sandercock, 
has been modified in several details to allow for stable operation and
high contrast.
The six-pass optics has been placed in a thermally isolating housing,
and the scanning stage is actively temperature stabilised.
The analogue-electronic stabilisation of the interferometer piezos
is replaced by computer control.
Entrance and exit pinholes are spatially separated from the six-pass optics,
so that the most critical alignements can be done without disturbing the
interferometer operation.
By placing additional masks in the six-pass optics,
in particular on mirror surfaces,
the cross-talk between different passes could be reduced by several orders,
and a straylight rejection of better than $10^{11}$ was achieved.

Depending on the free spectral range,
the instrument is used in series with an interference filter of
0.15 or 1\,GHz band width (Andover);
these filters are placed in a special housing
with active temperature stabilisation.
Furthermore, to account for long-term drift, 
the instrument function is redetermined periodically 
by automatic white-light scans.

Although the spectrometer guaranteed excellent straylight rejection,
special care was taken to prevent direct or diffuse reflections of laser light 
from entering the instrument.
Therefore, instead of 180$^\circ$ back-scattering,
a 169$^\circ$ VH geometry was chosen.
From the intensity 
transmitted through the ``ghosts'' of the tandem instrument,
we conclude that the straylight was about $10^5$ times weaker than
the inelastic scattering from PC,
and completely negligible 
compared to the detector's dark count rate of about 2.5\,sec$^{-1}$.
After subtraction of these dark counts,
and normalisation to the corresponding white light scans,
the resulting spectra showed excellent detailed balance symmetry.

The sample material was from the same source 
as for the neutron scattering experiment, 
and was vacuum sealed in a Duran cell.

\subsection{\boldmath Susceptibilities Around $\omega_\sigma$}

For a precise determination of the spectral lineshape,
subsequent measurements over different spectral ranges
were performed after stabilising the temperature over night.
The most restrictive determination of the exponent parameter was possible
at $T=216\,$K.
As shown in Fig.~\ref{Fl1su}, 
the matching of the three overlapping spectral ranges is excellent.
Except for the leaking VV Brillouin mode,
the VH susceptibility is described over more than two decades by the
mode coupling asymptote $g_\lambda$. 
Fits yield $\lambda=0.72\pm0.01$,
which is confirmed by measurements at other temperatures
as well as by a bulk of earlier experiments 
we had performed under less ideal experimental conditions.
This result is at the margin of the
error range in the literature value $0.78\pm0.05$ \cite{DuLC94};
the figure shows that 0.78 itself is clearly incompatible with our present data.

Secondly,
for two different spectral ranges we measured temperature series.
Fig.~\ref{FlTsu} shows some of the composite susceptibilities.
From fits with fixed $\lambda=0.72$,
the frequencies $\omega_\sigma$ were obtained.
Then, the individual susceptibilities, 
measured around interferometer mirror spacings $z_0=0.8$ and 2.4\,mm
were fitted with fixed $\lambda$ and $\omega_\sigma$
so that we obtained two independent data sets for the amplitudes $\chi_\sigma$.

The temperature dependence of $\omega_\sigma$ is shown in Fig.~\ref{Flpar}a.
Data between 190 and 230\,K extrapolate to the same $T_{\rm c}=182\pm1$\,K
as found from neutron scattering. 
Extending the fit range to higher temperatures leads to a $T_{\rm c}$ 
which is about 2\,K lower.
Again, these results are marginally compatible with $T_{\rm c}=187\pm5$\,K
from the earlier light scattering study \cite{DuLC94}.

Measuring amplitudes in light scattering is difficult,
and the prediction (\ref{Epars}) can be verified only
over a reduced temperature range.
Results are visualised best in a logarithmic plot of $\chi_\sigma$ 
{\it vs.} $T-T_{\rm c}$ (Fig.~\ref{Flpar}b)
which suggests there are two different regions 
in which $\chi_\sigma$ is proportional to $|\sigma|^{1/2}$,
separated by some step.
Without further experiments, we must leave open whether this step
comes from the sample or presents an experimental artifact,
due for instance to distortions of the optical paths.
We note, however, that after 30 hours and a full temperature cycle,
we were able to reproduce an amplitude $\chi_\sigma$ within less than 1\,\%.
Thus, irregularities in the amplitude are due not to a drift in time,
but mainly to reversible effects of temperature variation.

Putting aside the amplitude problem,
our results are in excellent accord with the neutron scattering data 
of sect.~II.
This motivates us
to reconsider dielectric loss data
which have been analysed previously with $\lambda=0.78$.

\section{Other Spectroscopies}

\subsection{Dielectric Loss}

Dielectric spectroscopy on PC has been described recently
 \cite{LuPD97b,LuPD96c,ScLB99}.
The measurements extended over a wide range of temperatures and many decades
in frequency,
and the analysis has addressed different issues
which are currently debated in the context of glass transition dynamics.

Within the GHz--THz range,
the dielectric susceptibility~$\epsilon''(\omega)$ passes through a minimum
as suggested by MC theory.
Fits with Eq.~(\ref{EminL}), approximated as a sum of two power laws \cite{x39},
gave $\lambda=0.78$,
and from the temperature dependence of frequency and amplitude (\ref{Epars})
$T_{\rm c}\simeq187\,$K was found.
Though these values had strong support in the existing literature,
they differ from our neutron and light scattering results.
It is therefore interesting to ask whether the dielectric loss would
also allow for $\lambda=0.72$.

Since the available dielectric data are even noisier  
than the neutron susceptibilities,
there is no unique way to determine the scaling range within which
Eq.~(\ref{EminL}) applies;
therefore, any value of $\lambda$ depends on the choice of the fit range.
In the previous analysis \cite{ScLB99},
fits were applied between about 1 and 600\,GHz.
For temperatures around 200\,K this range covered both sides
of the minimum equally well.
When $\lambda=0.72$ is imposed in an iterative master curve construction,
the scaling range evades towards lower frequencies,
as can be seen from the fits in Fig.~\ref{Fdie72}.

For direct comparison,
Fig.~\ref{Fd2ma} shows master curves constructed with 
two different values of $\lambda$.
In the upper curve, with $\lambda=0.78$,
the measured susceptibilities have been
rescaled with exactly the $\omega_\sigma$ and $\epsilon_\sigma$ 
shown in Fig.~8 of Ref.~\onlinecite{ScLB99}.
The rescaling works particularly well on the high-frequency side of the minimum
and shows nicely the $\omega^a$ limit which has remained elusive
in so many other experimental investigations.
The lower curve,
constructed with $\lambda=0.72$, %
shows that the scaling property is completely lost above $\omega_\sigma$,
whereas it has significantly improved in the low-frequency wing.
Furthermore,
the $\omega^{-b}$ limit of $g_\lambda$ describes the data
nearly up to the $\alpha$-relaxation maximum
where they almost coincide,
except for the lowest two temperatures.

This unexpected observation motivates a new master curve construction
which is quite common in the conventional analysis of $\alpha$ relaxation,
but which has never before been applied to MC $\beta$ relaxation:
in Fig.~\ref{Fama}, the dielectric data are shown with original amplitude,
rescaled only in frequency
by the $\alpha$ maximum frequency $\omega_{\rm max}$.
As usual for $\alpha$ relaxation
(and in accord with the second scaling law of MC theory)
the amplitude and the line shape of the $\alpha$ peak 
are temperature independent.
But here, 
the scaling behaviour extends far down into the high-frequency wing ---
in fact, it extends as far as the data have been measured,
except again the two lowest temperatures, 193 and 203\,K.

However, a master curve which extends from the maximum up to the minimum
of a dynamic susceptibility cannot simultaneously obey the first and the
second scaling law of MC: we expect
\begin{equation}
  \omega_{\rm max}/\omega_{\rm min}\propto {|\sigma|}^{1/2b}\,,
\end{equation}
and, even more elementary, $\epsilon_{\rm min}\propto{|\sigma|}^{1/2}$
whereas $\epsilon_{\rm max}$ remains constant.

Since there is no doubt about the scaling of $\alpha$ relaxation,
we are bound to conclude that the dielectric data do not reach 
the asymptotic regime of fast $\beta$ relaxation,
except in a rather small temperature range that extends
at best to about 210\,K.
For higher temperatures,
the high-frequency wing of $\alpha$ relaxation,
though technically describable by $g_\lambda$, 
does not represent the first scaling law limit.

This conclusion is independent of any fitting,
and does in particular not depend upon an imposed value of~$\lambda$.

\subsection{Solvation Dynamics}

The solvation response of $s$-tetrazine has been measured in PC
from 1.5 to 100\,ps \cite{MaBB95}.
A MC analysis in the time domain identified both $\alpha$- and $\beta$-scaling
regions in the dynamics \cite{MaBB96}.
A unified analysis of both regions was consistent with MC theory for
a range of $\lambda$--$T_{\rm c}$ pairs.
In the original publication, $\lambda=0.78$ was fixed,
in accord with the light scattering analysis of Du {\it et al.} \cite{DuLC94}.
However, this value of $\lambda$ yielded 
a cross-over temperature $T_{\rm c}=176\,$K 
substantially below the value obtained by Du {\it et al.} ($T_{\rm c}=187\,$K).

The solvation data have been reanalysed by the same method used
in Ref.~\cite{MaBB96}, but using $\lambda=0.72$.
The analysis of the $\alpha$-scaling region is unchanged.
Figure~\ref{Fholes} shows the new beta-scaling plot
and the fit to $g_\lambda(t/t_\sigma)$,
both of which are as good as in the previous analysis.
The inset shows a temperature scaling plot 
of $t_\sigma$ and $\tau_\alpha$.
As expected, the scaling law deteriorates at temperatures far above $T_{\rm c}$.
Depending on the range of temperatures fit,
acceptable values of $T_{\rm c}$ lie in the range 178--182\,K.
Thus, the solvation response is consistent 
with both the $\lambda$ and $T_{\rm c}$ obtained in Sects.~II and III
by neutron and light scattering.

\section{Comparison of Dynamic Observables}

\subsection{Direct Comparison of Susceptibilities}

Neutron scattering,
light scattering with or without proper bandpass,
dielectric loss:
each experiment, taken alone, 
seemed in full accord with the asymptotic predictions of MC theory.
Taken together, the situation becomes more complicated.

In the fast $\beta$ regime,
any observable that couples to density fluctuations
is expected to tend towards the same asymptotic limit~(\ref{EminL}).
For neutron scattering,
this prediction takes the form of a $q$,$\omega$ factorisation,
and is confirmed, in PC as in other materials, over a wide frequency range.
However, it breaks down completely when depolarised light scattering
or dielectric loss are included.

In Fig.~\ref{F3sus3},
dynamic susceptibilities from 
neutron scattering, light scattering \cite{x40}, and dielectric loss
are shown for direct comparison
on an arbitrary intensity scale, but in absolute frequency units.
The result is in flagrant contradiction to any factorisation property:
there is no one temperature for which the measured 
$\epsilon''(\omega)$ and $\chi''(\omega)$ fall together;
in particular, their minimum frequencies differ systematically.

On a first sight,
this outcome is a bad surprise,
and could make us doubt whether 
fits of individual data sets with the asymptotic laws 
(\ref{EminL}) and~(\ref{Epars}) are meaningful at all.
On closer examination,
the discrepancies between the three data sets can be traced back
to two major differences:
to the individually different temperature ranges 
within which the $\beta$ asymptote applies, 
and to a systematic shift in the frequency scale $\omega_\sigma$.

\subsection{\boldmath Temperature Range of $\beta$ Relaxation}

Within idealised MC theory, 
all dynamic observables converge towards the same scaling limit
(\ref{EminL}) and~(\ref{Epars}),
characterised by just one lineshape parameter~$\lambda$ and one
frequency scale~$\omega_\sigma$.
This universality, however, is restricted to the lowest-order asymptote
and does not imply a universal radius of convergence:
the next-to-leading-order corrections already depend 
on the microscopic coupling \cite{FrFG97b}
so that different observables may reach the asymptote
at different temperatures and frequencies. 

Nevertheless, for neutron and light scattering,
as well as for solvation response,
(\ref{EminL}) and~(\ref{Epars}) hold over rather large temperature intervals.
In neutron scattering, (\ref{Epars}) holds best between 220 and 260\,K;
at 210\,K, the susceptibility minimum approaches the instrumental
resolution function, and the signal becomes very weak:
these technical limitations prevent us from following $S(q,\omega)$
closer to $T_{\rm c}$.
For light scattering, Fig.~\ref{Flpar} confirms (\ref{Epars}) 
with reasonable precision up to nearly 1.5\,$T_{\rm c}$.

On the other hand, 
the $\alpha$ master curve in Fig.~\ref{Fama} suggested 
that the range of $\beta$ scaling is rather small for dielectric loss.
This is corroborated by Fig.~\ref{F3ws} in which the $\omega_\sigma$ 
from fits with fixed $\lambda=0.72$ are compiled.
For the lowest temperatures 193--213\,K,
the minima of $\epsilon''(\omega)$ coincide with the light scattering data, 
whereas for higher temperatures
the dielectric $\omega_\sigma$ cross over 
to a much steeper temperature dependence characteristic for $\alpha$ relaxation
(actually even steeper than the MC prediction 
${|\sigma|}^{1/2a+1/2b}$).

\subsection{\boldmath Frequency Scales $\omega_\sigma$}

As Fig.~\ref{F3ws} shows,
the $\omega_\sigma$ from neutron scattering fall about 35\,\%
below the $\omega_\sigma$ from light scattering.
The good agreement of $\omega_\sigma$ from neutron scattering 
with ${t_\sigma}^{-1}$ from solvation dynamics 
excludes the possibility that the discrepancy
between neutron and light scattering lies only in
multiple scattering or other technical shortcomings of neutron scattering.
{\it Ergo}, at least one of the two scattering techniques does not see
the true $\beta$ asymptote of MC theory.

On the other hand, 
the observed scaling,
the quality of the $g_\lambda$ fits, and
the accord of the $\lambda$ and $T_{\rm c}$ 
call for a MC interpretation,
and indicate a transient behaviour that is closely coupled
to the universal $\beta$ asymptote.

\section{Discussion}

\subsection{\boldmath Strength of $\alpha$ Relaxation}

Although there is no universal criterion for the validity of $\beta$ scaling,
an upper limit may be given:
it is plausible that the asymptote~(\ref{EminL}) 
which is based on an expansion around the susceptibility minimum
 will no longer apply when the minimum ceases to exist.
This happens when the height of the minimum, $\chi_\sigma$,
becomes comparable to the maximum of $\chi''(\omega)$ in the vibrational band.
This condition also determines the highest temperature for which
the proportionality~(\ref{Epars}) can hold.

On the other hand, the temperature range over which~(\ref{Epars}) holds
can be related to the strength of the $\alpha$ peak.
To explain this relation, we refer back to 
the $\alpha$ master plot in Fig.~\ref{Fama}.
Besides the dielectric $\epsilon''(\omega/\omega_{\rm max})$,
the figure also shows 
$\chi''_{\rm VH}(\omega/\omega_{\rm max})$ from depolarised light scattering.
For both data sets, 
the same frequency scale $\omega_{\rm max}$ has been applied,
as determined from the maximum position of the dielectric $\alpha$ peak.
The intensity scale of the light scattering data is arbitrary 
and has been adjusted by a global factor so that the maximum
of the microscopic excitations has about the same height as for the
dielectric data which are measured in absolute units.

Compared to the high-frequency maximum,
the $\alpha$ peak is about four times weaker in light scattering
than in dielectric loss.
In the high-frequency wing, at, say, $\omega\simeq30\omega_{\rm max}$,
the ratio is reduced to a factor of about 2.
This reduction is due to different positions of the $\alpha$ maxima,
and to different slopes of the wings.

At still higher frequencies, the susceptibilities cross over towards
the minimum.
For the lowest temperatures shown in the figure,
the ratio of~2 between dielectric and light scattering data
survives in this frequency range.
Thus, for a low-temperature value of~$\sigma$,
the $\beta$ amplitude~$\chi_\sigma$ for light scattering
is about two times smaller than for dielectric loss.

On the other,
from the condition stated above,
and because we have rescaled the susceptibilities to about 
the same values at high frequencies, 
the temperature range of $\beta$ scaling ends for both techniques
at about the same absolute value of $\chi_\sigma$.
Using the proportionality~$\chi_\sigma\propto{|\sigma|}^{1/2}$,
the dielectric data reach this limit after a 4 times smaller
 temperature change than the light scattering data do.

Thus, the strength of $\alpha$ relaxation can explain
why the $\beta$ scaling regime extends over only 30\,K in dielectric loss,
compared to 90\,K or more in light scattering.

\subsection{\boldmath Frequency Range of $\beta$ Relaxation}

One can imagine many experimental imperfections that distort 
spectra measured in a scattering experiment.
For instance, 
multiple scattering could overlay $S(q,\omega)$ with a convolution with
itself.
But any such distortions
would affect the wings of the susceptibility much more than
the region around the minimum.
In particular, convolutions lead to similar corrections
as intrinsic next-to-leading-order terms.
Therefore,
it is not easy to explain the discrepancy observed in the $\omega_\sigma$.

We note, however, that the $g_\lambda$ fits to the neutron scattering data
in Fig.~\ref{Fnsufi} extend up to frequencies between 200 and 400\,GHz,
whereas in light scattering, the fitted $g_\lambda$ already deviate from the
measured susceptibilities a little above 100\,GHz.
It is therefore quite possible
that our neutron scattering analysis sees a preasymptotic transient
rather than the true $g_\lambda$.

This would ressemble the situation in glycerol \cite{WuHL94},
where the $\chi''(\omega)$ allowed for the construction of a master curve 
and followed some MC predictions,
although the full $\beta$ asymptote was not reached.
Numeric solutions of a simple two-correlator model showed
how such a scenario can arise from a MC ansatz \cite{FrGM97}.
Similar fits, with one slave correlator for each observable,
also work for propylene carbonate \cite{GoVo00}.

\subsection{Current Theoretical Developments}

Some of the questions raised by our experiments are 
also addressed by recent numerical solutions of MC equations
and molecular-dynamics simulations.

A MC analysis of the hard-sphere liquid as a function of wavenumber
has demonstrated that corrections to scaling are of differing importance
for different observables.
Analytic expansions have shown that corrections to the $\alpha$ process are
of higher order than corrections to the $\beta$ asymptote:
this may explain why $\alpha$ scaling holds over a wider range
than the factorisation property of $\beta$ relaxation \cite{FrFG97b}.

In a molecular-dynamics simulation 
of a liquid made of rigid diatomic molecules \cite{KaKS98b},
orientational correlation functions have been analysed
for different angular momenta $l=0,1,2,\ldots$
All correlators were found to fulfil the MC factorisation --- 
with the pronounced exception of $l=1$ for which the position of the
susceptibility minimum is shifted by a factor of~10 [{\it loc.\ cit.}, Fig.~8].
The authors attribute this peculiar behaviour to $180^\circ$ jumps,
which is not necessarily the right explanation for similar behaviour 
in more complex fluids.
A scenario with an underlying type A transition
for odd $l$ has been proposed recently \cite{Sin99}.
It has been noted before that the dielectric response couples only weakly
to vibrational excitations \cite{LuPD96b},
and the peculiar strength of the $\alpha$ peak in the $l=1$ correlator
has also been found in an analytic extension of MC theory 
to non-spherical particles \cite{ScSc97}.

A decomposition according to angular momenta might also explain
the excellent accord between neutron scattering and solvation response
found in Fig.~\ref{F3ws}, 
since both techniques see $l=0$
whereas light scattering might be dominated by $l=2$ \cite{LeDG97}.

\section*{Acknowledgements}

We thank H.~Z.~Cummins, M.~Diehl, J.~K.~Kr\"uger, H.~Leyser,
J.~R.\ Sandercock, A.~P.\ Sokolov, and
J.~Wiedersich for invaluable advice in setting up our Fabry-Perot spectrometer.

We are grateful to W.~Petry for continuous support,
and we thank him as well as 
M.~Fuchs, W.~G\"otze, A.~P.~Singh and T.~Voigtmann for fruitful discussions.

We acknowledge financial aid by
the Bundes\-mini\-sterium f\"ur Bildung, Wissenschaft, Forschung und Technologie
through Verbundprojekte 03{\sc pe}4{\sc tum}9 and 
03{\sc lo}5{\sc au\footnotesize 2}8
and through contract no.\ 13{\sc n}6917,
by the Deutsche Forschungsgemeinschaft under grant no.\ {\sc Lo}264/8--1,
by the European Commission
through Human Capital and Mobility Program ERB CHGECT 920001,
and by the National Science Foundation under {\sc Che9809719}.

The Laboratoire L\'eon Brillouin is a laboratoire commun CEA -- CNRS.




\narrowtext

\begin{table}
\caption
{Exponent parameter~$\lambda$ and cross-over temperature~$T_{\rm c}$
of propylene carbonate as determined by different experimental techniques.}
\label{LitTab}
\tablenotetext[1]{This column indicates whether
the MC fit was based on the Debye-Waller factor $f_q$ (\ref{Esqrt}),
the $\alpha$ time scale $\tau_\alpha$ (\ref{Etaua}),
or the $\beta$ relaxation parameters $\omega_\sigma$ and $\chi_\sigma$
(\ref{Epars}). The subscript $\sigma_\pm$ indicates that $T<T_{\rm c}$ data
have also been used.}
\tablenotetext[2]{Used as input.}
\tablenotetext[3]{Determination of MC parameters judged unreliable
 by the original authors.}
\tablenotetext[4]{The value $179\pm2$\,K obtained 
in an alternative analysis using extended MC theory
is not directly comparable to the $T_{\rm c}$'s from idealised theory.}
\tablenotetext[5]{Transient hole burning on $s$-tetrazine 
 in propylene carbonate. Original analysis.} 
\tablenotetext[6]{Original data from Ref.~\cite{BoHu85}.}
\begin{tabular}[th]{lllll}
method & param.\tablenotemark[1] & $\lambda$ & $T_{\rm c}\;$(K) & reference \\
\hline
neutron scattering & $f_q$ & -- & 210 & 
 \protect\onlinecite{BoHo91} \\
Brillouin scattering & $f_q$ & -- & $270\pm5$ &
 \protect\onlinecite{ElBT92} \\
Brillouin scattering & $f_q$ & 
 --\tablenotemark[3] & 
 --\tablenotemark[3] &
 \protect\onlinecite{DuLC94} \\
viscosity\tablenotemark[6] & $\tau_\alpha$ & 0.70 & 196 &
  \protect\onlinecite{BoET90} \\
{\it same data} 
 & $\tau_\alpha$ & 0.78\tablenotemark[2] & $188\pm3$ &
  \protect\onlinecite{DuLC94} \\
neutron scattering & $\tau_\alpha$ & 0.70\tablenotemark[2]& 180/188 &
 \protect\onlinecite{BoET90,BoHo91} \\
dielectric loss & $\tau_\alpha$ & 0.78\tablenotemark[2] & 187\tablenotemark[2] &
  \protect\onlinecite{ScLB99} \\
VH light scattering & $\omega_{\sigma_\pm}$, $\chi_{\sigma_\pm}$ 
  & 0.78 & $187\pm5$\tablenotemark[4] & 
 \protect\onlinecite{DuLC94} \\
dielectric loss & $\omega_\sigma$, $\chi_\sigma$ & 0.78 & 187 &
  \protect\onlinecite{LuPD97b,LuPD96c,ScLB99} \\
solvation response\tablenotemark[5]
& $t_\sigma$, $\tau_\alpha$ & 0.78\tablenotemark[2] & 176 &
  \protect\onlinecite{MaBB96} \\
{\it same data} 
& $t_\sigma$, $\tau_\alpha$ & 0.72\tablenotemark[2] & $180\pm2$ & 
 {\it this work} \\
neutron scattering & $\omega_\sigma$, $\chi_\sigma$ & $0.72$ & 
 $182$ & {\it this work} \\
VH light scattering & $\omega_\sigma$, $\chi_\sigma$ 
  & 0.72 & $182$ & {\it this work} \\
\end{tabular}
\end{table}


\begin{figure}\caption 
{Incoherent neutron scattering law $S(q,\omega)$ of propylene carbonate 
at $T=207-210\,$K and $q=1.2\,$\AA$^{-1}$,
measured on two time-of-flight spectrometers 
under different experimental conditions:
Mib\'emol ($\lambda_{\rm i}=8.5\,$\AA, circular Al container) 
and {\sc Neat} ($\lambda_{\rm i}=5.5\,$\AA, 
linear arrangement of glass capillaries).
Figure (b) repeats the data of (a) on a double logarithmic scale
and compares them to the measured resolution functions (lines).
Above 20\,GHz, the quasielastic spectra agree within about statistical error.}
\label{Fnsqw}
\end{figure}

\begin{figure}\caption 
{Decomposition of $\chi''_q(\omega)$ 
according to the factorisation~(\protect\ref{Efact})
for Mib\'emol (210\,K) and {\sc Neat} (207\,K) data.
(a) The amplitude $h_q$ has been determined by least-squares matching 
between 50 and 2500\,GHz. 
Other methods yield almost identical results. ---
(b) The spectral function $\chi''(\omega)$
is obtained after dividing measured $\chi''_q(\omega)$ by $h_q$.
At low frequencies, the accord between both experiments is excellent;
however, at high wavenumbers and especially at high frequencies,
deviations are strong and systematic.
Multiple scattering 
and imperfect correction for energy-dependent detector efficiency 
are probably the main causes for distortions of the spectral line shape.}
\label{Ffact1}
\end{figure}

\begin{figure}\caption 
{Rescaled susceptibilities $\chi''_q(\omega)/h_q$.
The resolution-broadened elastic line has been cut off,
except at 285\,K where the $q$-dependent $\alpha$ relaxation 
can be fully resolved.
The factorisation (\ref{Efact}) is confirmed for both
the mode-coupling $\beta$ regime and the vibrational spectra in the THz region.}
\label{Fnsu}
\end{figure}

\begin{figure}\caption 
{Susceptibilities $\chi''(\omega)$, 
obtained as $q$-independent averages $\chi''_q(\omega)/h_q$
from Mib\'emol data, 
for temperatures varying in 10\,K steps from 210 to 260\,K.
Solid lines are fits 
with the mode coupling asymptote $g_\lambda(\omega/\omega_\sigma)$
(\protect\ref{EminL}) with $\lambda=0.72$ fixed.
The fit ranges, indicated by full symbols, 
have been determined in a self-consistent iteration 
from Fig.~\protect\ref{F3nma};
at low frequencies, they are limited by resolution or $\alpha$ relaxation,
at high frequencies by the low-lying vibrational peak. ---
In the inset, the logarithms of the fit parameters 
$\chi_\sigma$ and $\omega_\sigma$ are plotted against each other.
The straight line shows the slope $a=0.318$ expected from theory.}
\label{Fnsufi}
\end{figure}

\begin{figure}\caption 
{Susceptibility master curves 
$\chi''(\omega) / \chi_\sigma$ vs. $\omega/\omega_\sigma$ from the 
data set of Fig.~\protect\ref{Fnsufi}.
Fitting the same original data with different values of~$\lambda$ 
yields different scales $\chi_\sigma$ and $\omega_\sigma$
which lead then to distinctly differing master curves.
The temperature dependence~(\protect\ref{Epars})
of $\chi_\sigma$ ($\square$) 
and $\omega_\sigma$ ($\blacklozenge$) is tested in the insets.
---
(a) Iterative rescaling yields a $\lambda$ of about 0.69
(but unstable and depending much upon idiosyncrasies of the fitting procedure).
The  temperature dependence of $\chi_\sigma$ and $\omega_\sigma$ 
is in conflict with~(\protect\ref{Epars}). ---
(b) For $\lambda\simeq0.72$, 
the rescaling is as good as for 0.69,
and $\chi_\sigma$ and $\omega_\sigma$ extrapolate from 220--260\,K
to a consistent~$T_{\rm c}=182\,$K. ---
(c) With $\lambda=0.78$ as suggested by part of the literature,
the master curve is of poorer quality, 
fits with $g_\lambda$ show that the imposed value of $\lambda$ is
not self consistent,
and $\chi_\sigma$ and $\omega_\sigma$ are again 
in conflict with~(\protect\ref{Epars}).}
\label{F3nma}
\end{figure}

\begin{figure}\caption 
{Susceptibility $\chi_{\rm VH}(\omega)$ from depolarised light scattering
from propylene carbonate at 216\,K,
measured at the Fabry-Perot-Sandercock tandem interferometer at Garching,
with $\lambda_{\rm i}=514.5$\,nm and 
with mirror spacings $z_0=0.8,2.4$ and 7.2\,mm.
The matching of the three data sets is excellent and would extent 
even further without the VV Brillouin mode leaking around 10\,GHz.
The solid lines is a fit with the mode-coupling asymptote
$\chi_\sigma g_\lambda(\omega/\omega_\sigma)$ with $\lambda=0.72$.
The dotted line corresponds to the mean literature value $\lambda=0.78$.}
\label{Fl1su}
\end{figure}

\begin{figure}\caption 
{Depolarised light scattering susceptibilities,
combined from two spectral ranges ($z_0=0.8$ and 2.4\,mm),
as function of temperature.
Solid lines are mode-coupling fits with fixed $\lambda=0.72$.
Open symbols indicate regions which had to be excluded from the fits:
above about 80\,GHz the cross-over to the non-universal vibrational spectrum;
around 8--15\,GHz the Brillouin line, 
and at the highest temperature 
the beginning of $\alpha$-relaxation.}
\label{FlTsu}
\end{figure}

\begin{figure}\caption 
{Scaling parameters obtained 
from mode-coupling fits.
The rec\-ti\-fied plot (a) confirms that $\omega_\sigma$
evolves with ${|\sigma|}^{1/{2a}}$.
When fitting only the full symbols, $T_{\rm c}=182\pm1\,$K is obtained in
full accord with the neutron scattering result from 
Fig.~\protect\ref{F3nma}. When extending the temperature range,
$T_{\rm c}$ tends to decrease by a few K. ---
The amplitudes,
determined in\-de\-pen\-dent\-ly for two spectral ranges,
 are ad\-van\-ta\-geous\-ly represented (b) in a logarithmic plot
{\it vs.} $T-T_{\rm c}$ with $T_{\rm c}=182\,$K as determined above.
The solid line shows the slope 1/2 expected from theory.}
\label{Flpar}
\end{figure}

\begin{figure}\caption 
{Dielectric loss data as published in 
Refs.~\protect\cite{LuPD97b,LuPD96c,ScLB99}.
Solid lines are MC fits (\ref{EminL}) with fixed $\lambda=0.72$.
This figure can be compared directly 
to Fig.~7 in Refs.~\protect\cite{ScLB99}
where the same data are fitted with $\lambda=0.78$.}
\label{Fdie72}
\end{figure}

\begin{figure}\caption 
{Dielectric loss data \protect\cite{LuPD97b,LuPD96c,ScLB99},
rescaled to two versions of a master function 
$\epsilon''(\omega/\omega_\sigma)/\epsilon_\sigma$.
In the upper curve, the same $\omega_\sigma$ and $\epsilon_\sigma$
are used as determined in Ref.~\protect\onlinecite{ScLB99}
for $\lambda=0.78$.
For the lower curve, $\lambda=0.72$ was imposed which shifts the scaling
range towards the low-frequency wing of the susceptibility minimum.}
\label{Fd2ma}
\end{figure}

\begin{figure}\caption 
{In this master plot,
the frequency scale is taken from the position of the $\alpha$ relaxation peak
in dielectric loss, and applied to both 
dielectric loss $\epsilon''(\omega)$ 
(same data as in~\protect\cite{ScLB99})
and depolarised light scattering $\chi_{\rm VH}''(\omega)$ 
(scanned from Ref.~\protect\cite{DuLC94}).
While the $\epsilon''(\omega)$ are shown in absolute units,
the amplitude of $\chi_{\rm VH}''(\omega)$ is arbitrarily rescaled
by a global factor 
so that the microscopic peaks culminate at about equal height. ---
This scaling representation elucidates
how a strong $\alpha$ peak reduces the temperature range
for which a susceptibility minimum can be observed.}
\label{Fama} 
\end{figure}

\begin{figure}\caption 
{Solvation response of $s$-tetrazine in propylene carbonate as
measured by transient hole burning. 
Same data as in Fig.~6 of \protect\cite{MaBB96},
but now reduced with $\lambda=0.72$, $f_c=0.56$.
The solid curve is the scaling function $g_\lambda(t/t_\sigma)$.
The deviations at long times are attributed to $\alpha$ relaxation. ---
The inset ({\it cf.} Fig.~7 of \protect\cite{MaBB96})
shows the temperature scaling of $t_\sigma$ ($t_\beta$ in \protect\cite{MaBB96})
and $\tau_\alpha$ (both in ps),
using the exponents
$2a=0.636$ and $\gamma=2.395$ that correspond to $\lambda=0.72$.
The fits shown extrapolate to a common $T_{\rm c}=182\,$K in accord
with neutron scattering measurements.
(Good fits can be obtained over the range 178--182\,K).} 
\label{Fholes}
\end{figure}

\begin{figure}\caption 
{Direct comparison of susceptibilities 
from neutron scattering [Mib\'emol, this work],
depolarised light scattering \protect\cite{DuLC94},
and dielectric loss \protect\cite{LuPD97b,LuPD96c,ScLB99}.
The absolute scale is chosen arbitrarily (but temperature-independent)
to make the data approximately coincide in the low-temperature, 
low-frequency corner
where they possibly reach a common asymptotic regime.}
\label{F3sus3}
\end{figure}

\begin{figure}\caption 
{Frequencies $\omega_\sigma$,
determined from mode-coupling fits with the same $\lambda=0.72$,
for neutron scattering, light scattering,
dielectric loss, and solvation dynamics 
 (from the time-domain analysis converted to $\omega_\sigma = 1/t_\sigma$).
Temperatures are shown as $T-T_{\rm c}$ with $T_{\rm c}=182\,$K.
The solid line shows the slope $1/2a=1.572$ expected for the given value 
of~$\lambda$;
the dotted line, fitted to some dielectric data points,
has a slope of~3.1 which is even steeper than $1/2a+1/2b=2.4$ expected
for the $\alpha$-relaxation limit.}
\label{F3ws}
\end{figure}

\end{multicols}
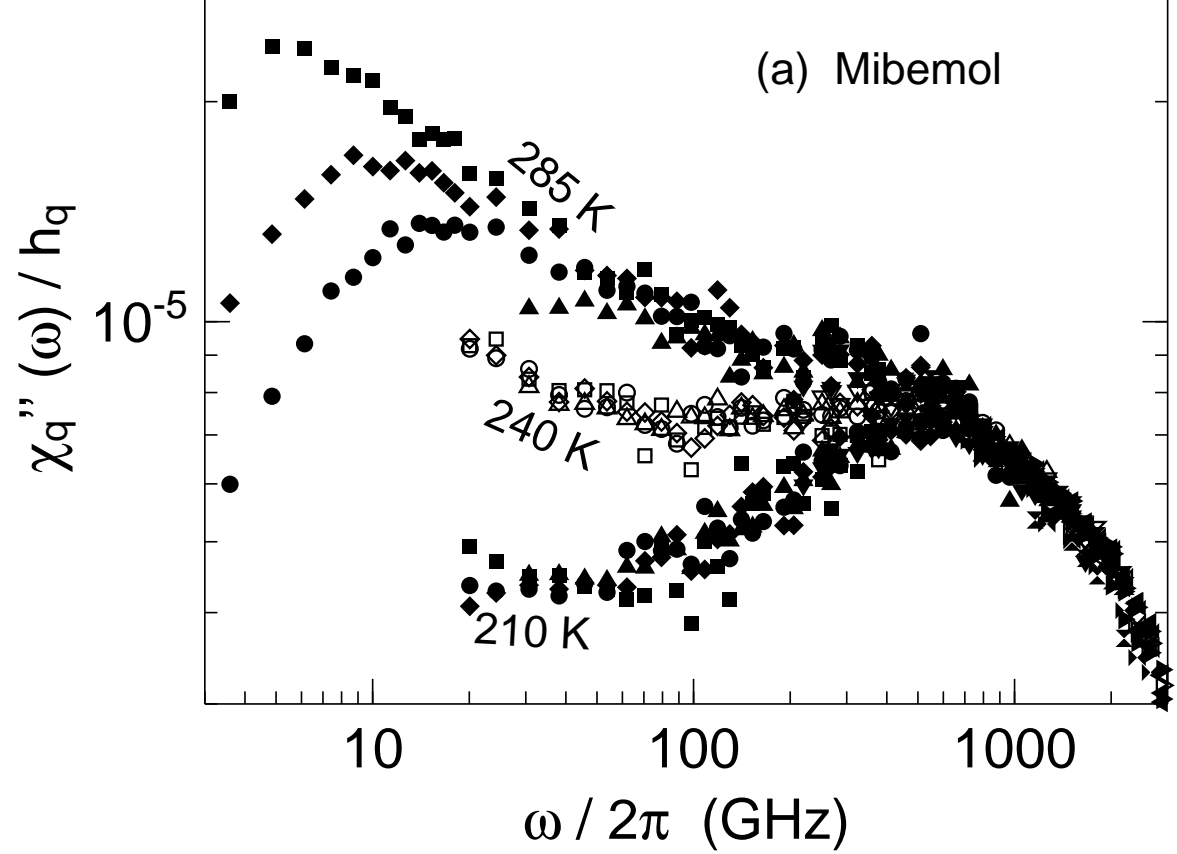
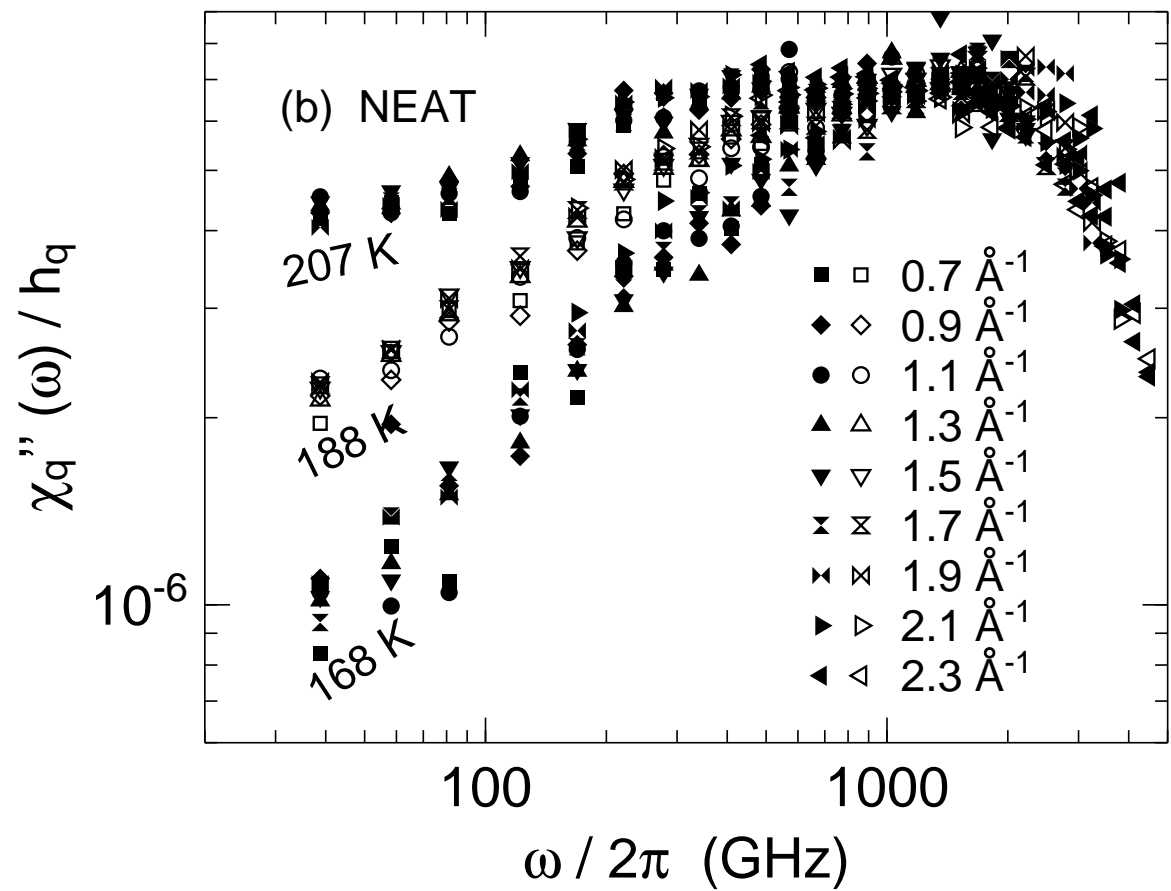
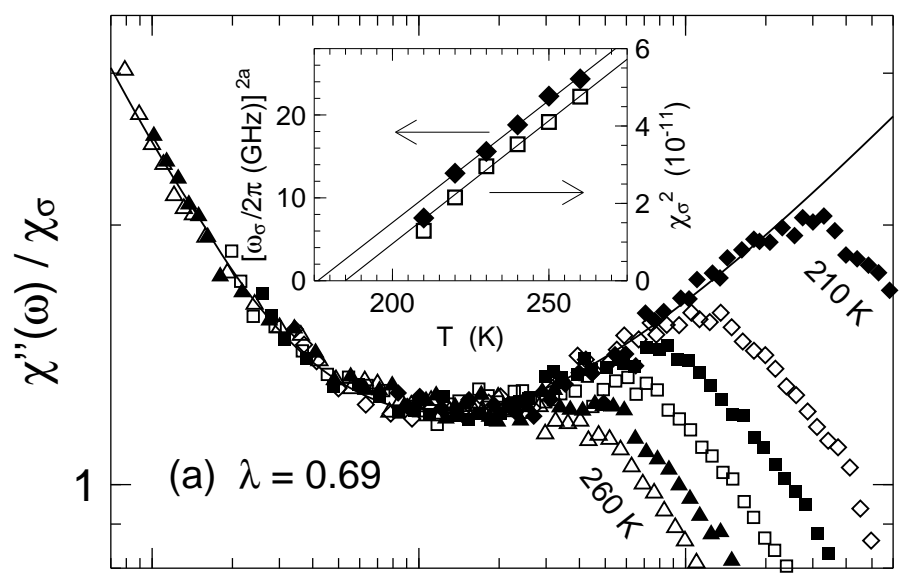
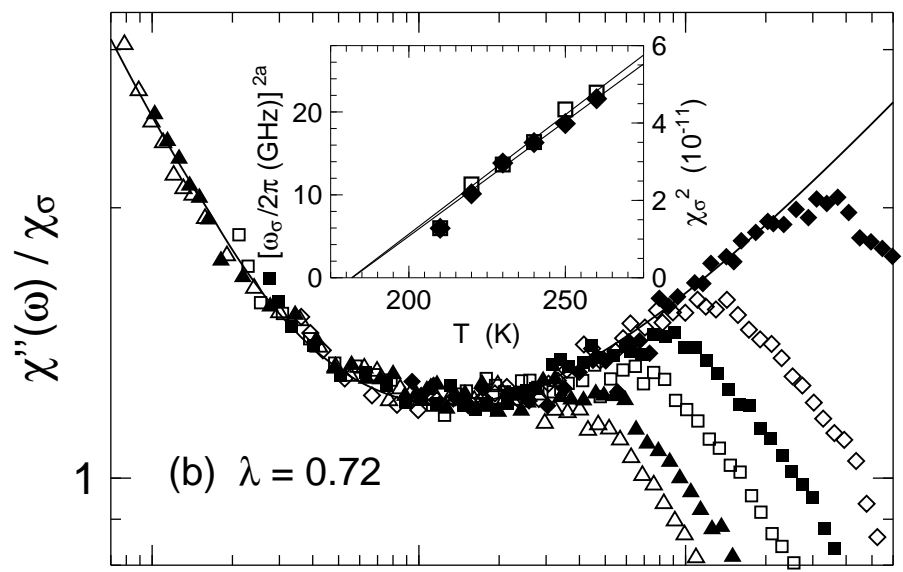
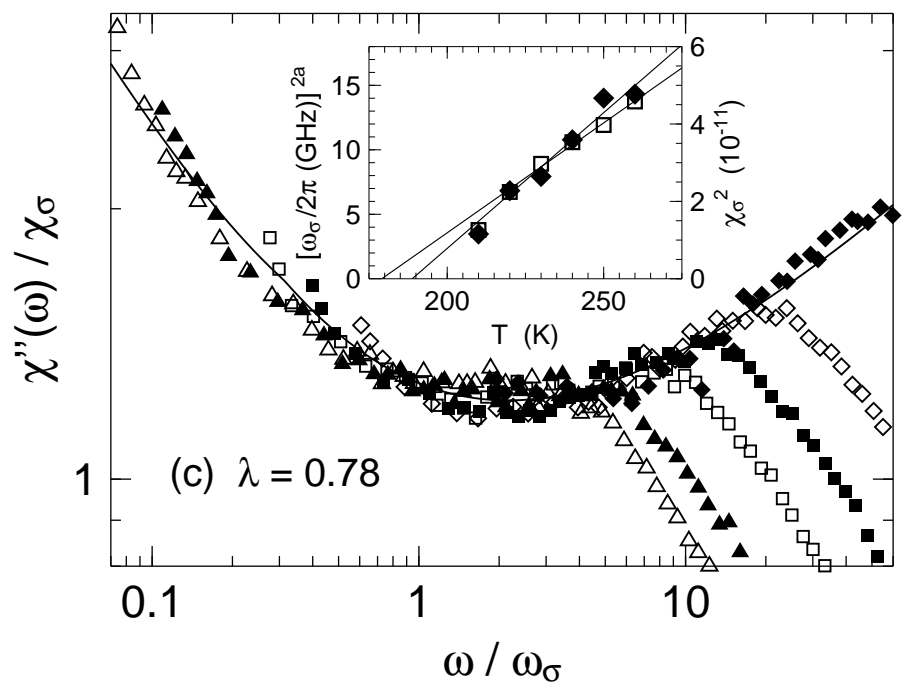
\end{document}